\documentclass[%
aip,
jap,
onecolumn,
amsfonts,
amsmath,
amssymb,
superscriptaddress
]{revtex4-1}%

\usepackage{graphicx}
\usepackage{subfigure}
\usepackage{dcolumn}
\usepackage{bm}

\usepackage{SIunits}

\begin{document}

\title{Black phosphorus-based anisotropic absorption structure in the mid-infrared}

\author{Tingting Liu}
\affiliation{Laboratory of Millimeter Wave and Terahertz Technology, School of Physics and Electronics Information, Hubei University of Education, Wuhan 430205, China}

\author{Xiaoyun Jiang}
\affiliation{Wuhan National Laboratory for Optoelectronics, Huazhong University of Science and Technology, Wuhan 430074, China}

\author{Chaobiao Zhou}
\affiliation{Wuhan National Laboratory for Optoelectronics, Huazhong University of Science and Technology, Wuhan 430074, China}
\affiliation{College of Mechanical and electrical engineering, Guizhou Minzu University, Guiyang 550025, China}

\author{Shuyuan Xiao}
\email{syxiao@ncu.edu.cn}
\affiliation{Institute for Advanced Study, Nanchang University, Nanchang 330031, China}

\date{\today}

\begin{abstract}
Black phosphorus (BP), an emerging two-dimensional (2D) material with intriguing optical properties, forms a promising building block in optics and photonics devices. In this work, we propose a simple structure composed of BP sandwiched by polymer and dielectric materials with low index contrast, and numerically demonstrate the perfect absorption mechanism via the critical coupling of guided resonances in the mid-infrared. Due to the inherent in-plane anisotropic feature of BP, the proposed structure exhibits highly polarization-dependent absorption characteristics, i.e., the optical absorption of the structure reaches 99.9$\%$ for TM polarization and only 3.2$\%$ for TE polarization at the same wavelength. Furthermore, the absorption peak and resonance wavelength can be flexibly tuned by adjusting the electron doping of BP, the geometrical parameters of the structure and the incident angles of light. With high efficiency absorption, the remarkable anisotropy, flexible tunability and easy-to-fabricate advantages, the proposed structure shows promising prospects in the design of polarization-selective and tunable high-performance devices in the mid-infrared, such as polarizers, modulators and photodetectors.
\end{abstract}

\maketitle

\section{\label{sec:1}Introduction}

As a new family of nanomaterials, two-dimensional (2D) materials with atomic-scale thickness exhibit advantages of robustness, mechanical flexibility and easy integration over traditional bulk material, and thus provide unprecedented possibilities in developing electronic and optoelectronic devices with desired physical properties. Graphene is the most popular member among 2D materials, and possesses the highest carrier mobility ($\sim$20000 cm$^{2}$/V$\cdot$s), but the zero-bandgap nature sets the obstacle in the applications where high on-off ratio is required.\cite{novoselov2004electric, bonaccorso2010graphene} Due to the thickness-dependent band gap, black phosphorus (BP) thin film is brought to the spotlight since its exfoliation from its bulk crystal in 2014.\cite{li2014black} In bulk form, BP has a band gap of 0.3 eV which is expected to increase as the thickness of BP thin film decreases and finally reach 2 eV for monolayer form.\cite{xia2014rediscovering, tran2014layer, zhang2015broadband} Moreover, the band gap of BP can be efficiently tuned via electrical gating approach, offering additional degrees of freedom to design tunable compact devices.\cite{low2014tunable, roldan2017black, deng2017efficient} Another distinct feature of BP is the high in-plane anisotropy stemming from the puckered structure of phosphorus atoms, which enables the development of novel polarization dependent optoelectronic devices.\cite{qiao2014high, wang2015highly} These attractive properties make BP an excellent material in the infrared regime. However, as the consequence of the intrinsic atomic thickness, the inherent optical absorption of BP-based devices is usually quite low, severely limiting the efficiency of these devices.

Inspired by the advancements in graphene plasmons with field confinement merit and tunable benefit in the spectral range spanning from infrared to terahertz range, great efforts have been devoted to discovering and exploring the plasmonic resonances of BP in different structures or systems.\cite{low2014plasmons, liu2016localized, xiong2017strong, ni2017surface, lu2017strong} However, most of the BP-based resonant structures exhibit weak light absorption with low doping concentration and low sensitivity. An alternative way for absorption enhancement is to introduce the multilayer structure of BP into the design.\cite{wang2017dual, cai2019anisotropic, xiao2019tunable} Nevertheless, the relatively complicated fabrication techniques are required for the high absorption rate in the multilayered structure. In addition, the theory of coherent absorption has also been demonstrated for absorption enhancement of BP, while the additional configurations is necessary for the proper phase modulation of two coherent incident waves.\cite{wang2018tunable, guo2019broadband} Most recently, a total absorption scheme in 2D materials based on critical coupled mode theory has attracted extensive attentions.\cite{piper2014total, jiang2017tunable, fan2017monolayer, li2018wavelength, akhavan2018narrow, qing2018tailoring} Especially, Qing et al. presented a perfect absorber by critical coupling of BP with guided resonances of a photonic crystal slab and demonstrated the sensitivity of the absorption to the slab parameters. Given the fabrication processes of the absorption structure composed of the photonic crystal slab with high index contrast are complicated in practice, the design of a simple BP-based structure with high absorption, flexible tunability and easy fabrication process is still required.

In this work, we propose a BP-based perfect absorption structure via critical coupling in the mid-infrared, where BP is coupled to a 2D periodic polymer structure with low index contrast. This simple structure not only efficiently enhances the light-BP interaction with the optical absorption up to 99.9$\%$, but also demonstrates the desirable capability of tailoring the absorption by controlling the conditions of critical coupling, including electron density of BP, geometrical parameters and incident angle of light. Furthermore, the BP-based structure exhibits distinct absorption characteristics under TM- and TE-polarized illumination due to the in-plane anisotropy of BP. Thus, this proposed structure shows great potentials for the high-performance BP-based optics and photonics devices in the mid-infrared.

\section{\label{sec:2}Structure and model}

The schematic diagram of the proposed absorption structure is illustrated in Fig. 1(a). A monolayer BP is sandwiched between a Polydimethylsiloxane (PDMS) layer with a periodic arrays of circular air holes and a MgF$_{2}$ layer, and a gold (Au) layer is deposited on the back of the MgF$_{2}$ layer to prevent the transmission of incident light. The side-view of the structure and the geometrical parameters are displayed in Fig. 1(b). The refractive indices of PDMS, air and MgF$_{2}$ are taken to be 1.37, 1 and 1.34, respectively.\cite{dodge1984refractive, querry1987optical} The permittivity of Au is given by the Drude model.\cite{ordal1985optical} The equivalent relative permittivity of BP can be derived by the surface conductivity which is dependent upon incident polarization due to the different electron mass along armchair and zigzag direction.\cite{liu2016localized, xiong2017strong} Fig. 1(c) and (d) displays the polarization-dependent permittivity of BP, and demonstrates its anisotropic optical behaviors.
\begin{figure}[htbp]
\centering
\includegraphics
[scale=0.60]{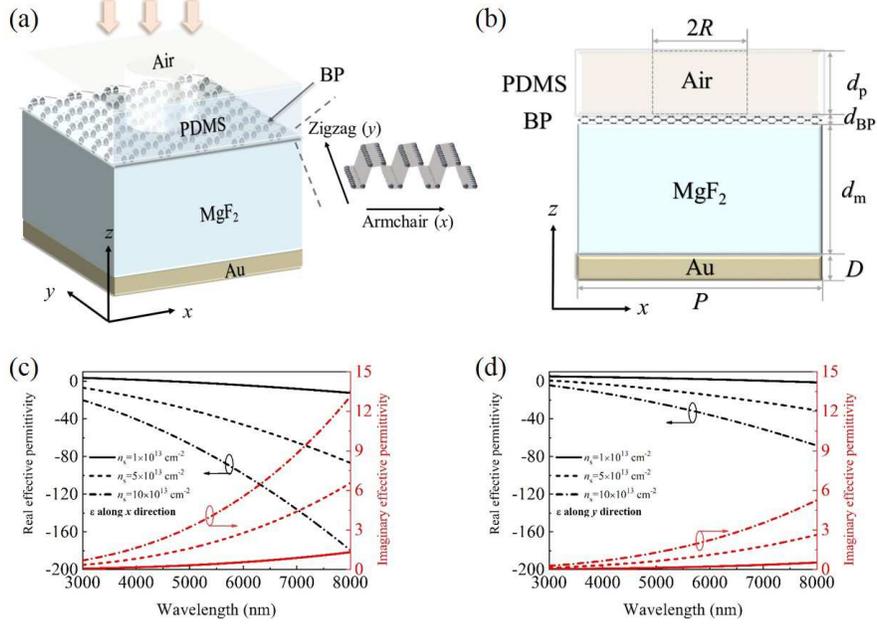}
\caption{\label{fig:1} (a) Schematic diagram and (b) the side-view of the proposed absorption structure. The inset illustrates the structure of monolayer BP. The lattice period in both $x$ and $y$ direction is denoted as $P$. The thickness of Au layer, MgF$_{2}$ layer, BP layer and PDMS layer are represented by $D$, $d_{m}$, $d_{BP}$ and $d_{p}$, respectively. The radius of the air holes is $R$. The effective permittivity of BP along (c) $x$ direction and (d) $y$ direction. The black and red lines denote the real and imaginary parts, respectively.}
\end{figure}

The numerical simulations for the absorption characteristics of the proposed structure are carried out using finite difference time domain (FDTD) method. The periodic boundary conditions are applied in both $x$ and $y$ direction and the plane waves are normally incident from $-z$ direction while the perfectly matched layer are employed to absorb all the light out of the boundaries along the propagation direction. The non-uniform mesh is adopted, and the minimum mesh size inside the BP layer equals 0.25 nm and gradually increases outside the BP layer to balance storage space and computing time. The absorption of the structure can be calculated from $A=1-T-R$ where $T$ and $R$ are transmission and reflection, respectively. Because of the Au layer as the metallic mirror to block the transmission, the absorption can be finally simplified as $A=1-R$.

In the proposed configuration, the concept of critical coupling is employed for perfect absorption via coupling the guided resonance to the lossy BP. The in-plane periodicity in the PDMS layer consisting of square lattices of air holes enables phase-matched coupling between the guided mode and free-space radiation, leading to the guided resonance with electromagnetic field significantly confined within the structure. Hence, the incident light can be coupled with the guided resonance and the absorption would be remarkably enhanced in the vicinity of the resonance frequency. The coupled mode theory (CMT) can account for the phenomenon of critical coupling for optical absorption enhancement. The reflection coefficient in the coupled system is described as\cite{piper2014total, jiang2017tunable}
\begin{equation}
\label{eq:1}
\Gamma=\frac{y}{u}=\frac{i(\omega-\omega_{0})+\delta-\gamma_{e}}{i(\omega-\omega_{0})+\delta+\gamma_{e}},
\end{equation}
and the absorption is calculated as 
\begin{equation}
\label{eq:2}
A=1-|\Gamma|^{2}=\frac{4\delta\gamma_{e}}{(\omega-\omega_{0})^{2}+(\delta+\gamma_{e})^{2}},
\end{equation}
where $\omega_{0}$ is resonant frequency, $\delta$ is the intrinsic loss and $\gamma_{e}$ is the external leakage rate. From the above equation, it can be seen that a 100$\%$ perfect absorption of the system would be realized at the resonance frequency $\omega_{0}$ when the intrinsic loss rate of the structure is the same with the external leakage rates of the guided resonances, i.e. $\delta=\gamma_{e}$. 

Under the critical coupling condition, the impedance of the proposed structure is supposed to match with that of the free space, i.e. $Z=Z_{0}=1$. For the proposed one-port structure, the effective impedance can be written as\cite{smith2005electromagnetic, szabo2010unique}
\begin{equation}
\label{eq:3}
Z=\frac{(T_{22}-T_{11})\pm\sqrt{(T_{22}-T_{11})^{2}+4T_{12}T_{21}}}{2T_{21}}.
\end{equation}
The two roots in this equation correspond to the two paths of light propagation, and the plus sign is taken to represent the positive direction here. Meanwhile, $T_{11}$, $T_{12}$, $T_{21}$ and $T_{22}$ are the elements of the transfer (T) matrix of the structure and their values can be calculated from the scattering (S) matrix elements as following,
\begin{equation}
\label{eq:4}
T_{11}=\frac{(1+S_{11})(1-S_{22})+S_{21}S_{12}}{2S_{21}},
\end{equation}
\begin{equation}
\label{eq:5}
T_{12}=\frac{(1+S_{11})(1+S_{22})-S_{21}S_{12}}{2S_{21}},
\end{equation}
\begin{equation}
\label{eq:6}
T_{21}=\frac{(1-S_{11})(1-S_{22})-S_{21}S_{12}}{2S_{21}},
\end{equation}
\begin{equation}
\label{eq:7}
T_{22}=\frac{(1-S_{11})(1+S_{22})+S_{21}S_{12}}{2S_{21}}.
\end{equation}

\section{\label{sec:3}Results and discussions}

To investigate the absorption characteristics of the structure, the geometrical parameters are optimized to satisfy the critical coupling conditions, and the parameter values are initially set as follows: the Au layer thickness $D$=200 nm, the MgF$_{2}$ layer thickness $d_{m}$=1500 nm , the monolayer BP thickness is $d_{BP}$=1 nm, the PDMS thickness is $d_{p}$= 700 nm, and the radius of air holes is $R$=700 nm. Here we adapt the electron doping of BP $n_{s}$=3$\times$10$^{13}$ cm$^{-2}$ accessible in experiments. The absorption spectra of the proposed structure are numerically simulated under the normally incident light, and the simulated absorption spectra are depicted in Fig. 2(a) and (b). The absorption spectra for TM and TE polarizations exhibit significant direction dependent properties resulting from the inherent in-plane anisotropic feature of BP. For TM polarization, the absorption spectrum with a perfect absorption up to 99.9$\%$ is obtained at the resonance wavelength of 4467.5 nm, while absorption spectrum for the case of TE polarization has the peak with 95.8$\%$ at 4476.5 nm. By comparison, the resonant wavelength as well as the resonance position for TE polarization shows a clear redshift, especially, the absorption is as low as 3.17$\%$ at 4467.5 nm, i.e., the resonance wavelength of TM polarization.

Here we focus on the investigation of absorption spectrum in the case of TM polarization due to the perfect absorption merit. The full width at half maximum (FWHM) is $\Delta\lambda$=3.96 nm, indicating the extremely narrow bandwidth for spectral selective absorption. The quality factor $Q$ which can be defined as $Q=\lambda_0/\Delta\lambda$ reaches the value of 1128.17. Meanwhile, the theoretical absorption spectrum based on CMT is also depicted in Fig. 2(a) which exhibits an excellent agreement with the simulation results. According to CMT, the loss and external leakage of the structure is $\delta=\gamma_{e}=9.34\times10^{10}$ Hz. Then the theoretical quality factor $Q$ is calculated as 1128.8 by $Q_{CMT}=Q_{\delta}Q_{\gamma}/(Q_{\delta}+Q_{\gamma})$ where intrinsic loss is defined as $Q_{\delta}=\omega_{0}/(2\delta)$ and the external leakage defined as $Q_{\gamma}=\omega_{0}/(2\gamma_{e})$. The nearly identical values of the theoretical $Q_{CMT}$ and the simulated $Q$ demonstrate the perfect absorption of the structure is attributed to the critical coupling. In addition, we calculate the effective impedance of the system in the vicinity of resonant wavelength, as shown in the inset of Fig. 2(a). It is clearly observed that the impedance of the structure is $Z=1.00+i8.03\times10^{-4}$ at resonance, substantially equaling to that of the free space. Under this condition, the reflection of the structure is maximally suppressed because of the impedance matching with free space and the transmission is blocked by the metallic mirror, giving rise to the perfect absorption. To gain a deeper insight, Fig. 2(c) provides the distributions of electric field intensity $|E|$ at the resonance wavelength. When the structure is excited to the resonant state, the electric field is confined as the guided modes near the BP layer, leading to a remarkable absorption enhancement and the final perfect absorption of the structure.
\begin{figure}[htbp]
\centering
\includegraphics
[scale=0.52]{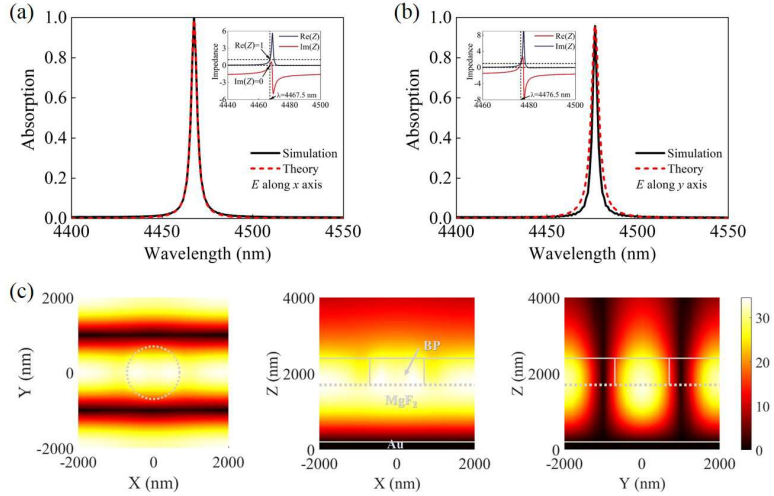}
\caption{\label{fig:2} Simulated and theoretical absorption spectra for (a) TM and (b) TE polarization. The insets are the effective impedances of the corresponding absorption spectra in the vicinity of the resonance wavelengths. (c) Distributions of electric field intensity $|E|$ at the resonance wavelength for TM polarization, including $x-y$ cross-section plane above the BP layer (left), along the $x-z$ (middle) and $y-z$ cross-section planes (right).}
\end{figure}

It is well known that the critical coupling depends on the match of the absorption rate of the structure and the leakage rate of the guided resonance, which gives us the clue to control the absorption characteristics by adjusting the two aspects. In the wavelengths of interest, the intrinsic absorption of the structure mainly contributed from the lossy BP layer. With tunable electron doping advantage, effective permittivity of BP can be dynamically controlled and then the critical coupling conditions as well as the absorption characteristics of the structure can be adjusted. Fig. 3(a) and (b) illustrates the variations of the absorption peak and resonance wavelength at different electron doping of BP under TM and TE polarizations, respectively. Obviously, the resonance wavelengths for TM and TE polarization both show a slight blue shift as the electron doping increases, which can be explained by the fact that the real part of the effective permittivity of BP becomes small with larger electron doping. It is also found that the absorption peak for TM polarization gradually increases from 94.8$\%$ with $n_{s}$=1$\times$10$^{13}$ cm$^{-2}$ to the perfect absorption of 99.9$\%$ with $n_{s}$=3$\times$10$^{13}$ cm$^{-2}$, then gradually decreases to 87.1$\%$ with $n_{s}$=11$\times$10$^{13}$ cm$^{-2}$. This can be explained by the match between the intrinsic absorption and the leakage rate of the structure. During the variations, the leakage rate of the structure can be considered as unchanged, $\gamma_e$= 9.34$\times10^{10}$ Hz, since only the electron doping $n_{s}$ of BP layer is changed here. Meanwhile, the other aspect, i.e., the intrinsic loss of the structure would gradually increases because the imaginary part of effective permittivity of BP becomes high as the electron doping increases. For more details of the increasing intrinsic loss $\delta$ and the steady leakage rate $\gamma_e$, the $\delta$ (5.6$\times10^{10}$ Hz) is smaller than $\gamma_e$ for $n_{s}$=1$\times$10$^{13}$ cm$^{-2}$, then equals to $\gamma_e$ for $n_{s}$=3$\times$10$^{13}$ cm$^{-2}$, and becomes larger (2.3$\times10^{11}$ Hz) than $\gamma_e$ for $n_{s}$=11$\times$10$^{13}$ cm$^{-2}$. It concludes that the system evolves from the states of over coupling to critical coupling and then under coupling as the electron doping increases. The above analysis also applies to TE polarization and the system goes through the states of over coupling to critical coupling for increasing electron doping $n_{s}$ from 1$\times$10$^{13}$ cm$^{-2}$ to 7.4 $\times$10$^{13}$ cm$^{-2}$, and then under coupling as $n_{s}$ further increases. In addition, the slight variation of absorption peak with above 90$\%$ amplitudes in a relatively broad electron doping region of BP reveals the robustness of the absorption structure, while the shift of resonance wavelength indicates the feasibility in a broad absorption spectrum regime.
\begin{figure}[htbp]
\centering
\includegraphics
[scale=0.55]{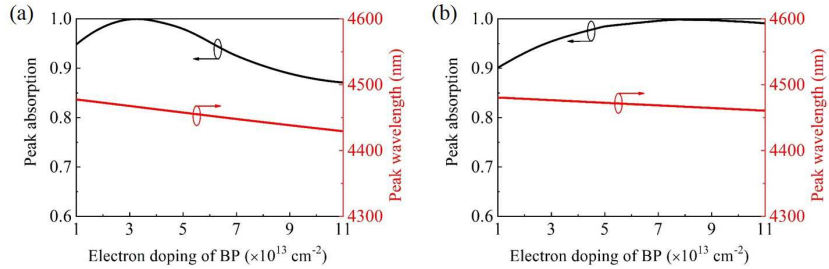}
\caption{\label{fig:3} Simulated absorption spectra at different electron doping $n_{s}$ of BP for (a) TM and (b) TE polarization.}
\end{figure}

On the other hand, we concentrate on the control of the external leakage rate of the structure when the intrinsic loss rate from BP keeps unchanged with electric doping $n_{s}$=3$\times$10$^{13}$ cm$^{-2}$. The influence of the geometrical parameters on the absorption of the whole structure is investigated via changing the air hole radius, the thickness of PDMS layer and MgF$_{2}$ layer, respectively. Here we focus on the investigation of absorption spectrum in the case of TM polarization. Fig. 4(a) illustrates the dependence of the absorption peak and the resonant wavelength on the air hole radius with other parameters are fixed. When the radius of air holes increases from 500 nm to 900 nm, the external leakage rate increases and the structure goes through the states of undercoupling, critical coupling and overcoupling in this radius range. Accordingly, the absorption peak of the structure firstly increases from 85.6$\%$ to 99.9$\%$ and then decreases to 90.2$\%$ during the modulation. At the same time, the wavelength of guided resonance i.e. absorption peak position shows a blue shift from 4492.5 nm to 4443.9 nm, which is attributed to the reduced effective refractive index of the guided resonance as air hole expands. Thus with a proper engineering of the air hole radius, the critical coupling conditions for absorption characteristics can be flexibly tuned.

The dependence of the absorption characteristics in the structure on the thickness of PDMS layer and MgF$_{2}$ layer is also considered in Fig. 4(b) and (c). When the PDMS layer thickness increases from 500 nm to 900 nm with the radius of air hole is fixed as $R=700$ nm, the absorption peak of the structure rises up from 92.3$\%$ to 99.9$\%$ and then goes down to 87.3$\%$. Compared to that, the absorption peak shows a drastically increase and then a sharp fall as the thickness of MgF$_{2}$ layer increases from 1300 nm to 1700 nm. This can be concluded that the external leakage rate of the structure is more sensitive to the variation of the MgF$_{2}$ layer thickness relative to the PDMS layer thickness. During their variations, the resonance wavelengths in Fig. 4(b) and (c) both exhibit redshift tendency with nearly linear variation, because the effective refractive index of the structure increases with the increasing thicknesses. Moreover, simulation results reveal that nearly perfect absorption up to 99$\%$ could be maintained with the air hole radius within the range from 650 to 720 nm, the PDMA layer thickness between 670 nm to 730 nm and the MgF$_{2}$ layer thickness between 1480 nm to 1530 nm when other parameters have fixed values. Therefore, the proposed structure not only realizes the high efficiency optical absorption with highly tunability, but also exhibits relatively large fabrication tolerances.
\begin{figure}[htbp]
\centering
\includegraphics
[scale=0.55]{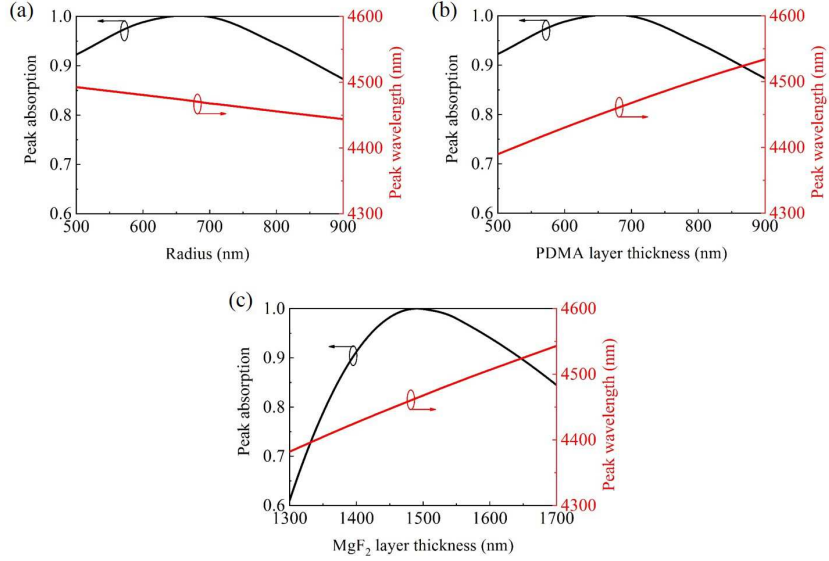}
\caption{\label{fig:4} Dependence of peak absorption and peak wavelength on geometrical parameters, (a) the air hole radius, (b)the thickness of PDMS layer and (c) the thickness of MgF$_{2}$ layer for TM polarization.}
\end{figure}

Finally, the absorption characteristics of the proposed structure under the oblique incidence is also investigated. Fig. 5(a) and (b) show the absorption as the function of incident angle and wavelength for TM and TE polarization, respectively. When the incident angle varies from 0 to 6$^{\circ}$, the absorption peak of the structure for TM polarization keeps nearly invariant and the resonance wavelength shows a slight blue shift. The absorption performance for TM polarization is insensitive to the incident angle within this range, and has potentials in designing integrated optoelectronic devices. For TE polarization, the phenomenon of wavelength splitting is observed and two major absorption peaks appear in the absorption spectra when the incident light is oblique. The two absorption peaks originate from the excitation of the guided resonance. In contrast with the stable absorption in TM polarization, the two absorption peaks exhibit distinct changes in resonance wavelength and absorption peak, showing promising applications in multispectral light detection. 
\begin{figure}[htbp]
\centering
\includegraphics
[scale=0.50]{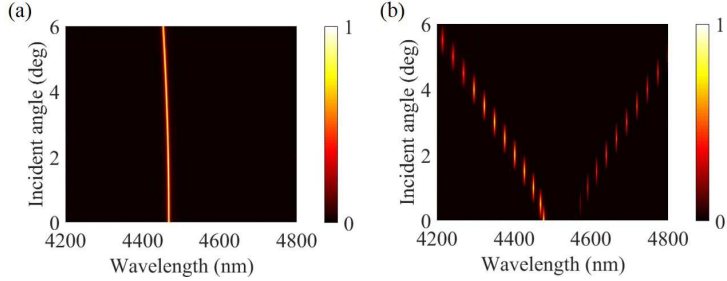}
\caption{\label{fig:5}  Dependence of absorption of the structure as the function of incident angle and wavelength for (a) TM polarization and (b) TE polarization. }
\end{figure}

\section{\label{sec:4}Conclusions}
In conclusions, a BP-based anisotropic perfect absorption structure is proposed and theoretically investigated in the mid-infrared. The perfect absorption up to 99.9$\%$ is attributed to the critical coupling of BP to the guided resonance in a periodic polymer lattice with low index contrast. The polarization-dependent absorption characteristics of the structure is obtained due to the anisotropic properties of BP. Moreover, the absorption peak and resonance wavelength can be flexibly tuned by adjusting the electron doping of BP, the geometrical parameters of the structure and the incident angles of light. In this work, the concerned mechanism of critical coupling for enhanced absorption is universal and the proposed structure shows advantages of high efficiency absorption, the remarkable anisotropy, flexible tunability, and easy-to-fabricate characteristics. Therefore, our results not only provide a good way to improve light-matter interaction for 2D materials, but also will play a significant role in the design of advanced BP-based devices.

\begin{acknowledgments}
This work is supported by the National Natural Science Foundation of China (Grant No. 61775064, 11847100 and 11847132), the Fundamental Research Funds for the Central Universities (HUST: 2016YXMS024) and the Natural Science Foundation of Hubei Province (Grant No. 2015CFB398 and 2015CFB502).
\end{acknowledgments}

\end{document}